\documentclass[12pt]{iopart}
\usepackage{iopams}  
\usepackage{dsfont} 
\usepackage{times} 
\usepackage{graphicx} 
\newcommand{\boldvec}[1]{\mbox{\boldmath$#1$}}

\newcommand{\annhilate}[2]{\hat{#1}^{\vphantom\dag}_{{#2}}}
\newcommand{\create}[2]{\hat{#1}^{\dag}_{{#2}}}

\newcommand{\eq}[1]{Eq.~(\ref{#1})}
\newcommand{\bra}[1]{\langle#1|}
\newcommand{\ket}[1]{|#1\rangle}

\newcommand{\scp}[2]{\langle #1\!\mid\!#2\rangle}

\newcommand{\ie}{i.\thinspace{}e., }

\newcommand{\bx}{{\mathbf{x}}}
\newcommand{\bp}{{\mathbf{p}}}

\newcommand{\text}[1]{{\rm #1}}

\begin{document}
\paper{Resonant Feshbach scattering of fermions in one-dimensional optical
  lattices} \author{M.~Grupp, R.~Walser, W.\,P.~Schleich} \address{ Institut
  f{\"u}r Quantenphysik, Universit{\"a}t Ulm, Albert-Einstein-Allee 11,
  D-89081 Ulm, Germany} \ead{michael.grupp@uni-ulm.de} \author{A.~Muramatsu}
\address{ Institut f\"ur Theoretische Physik III, Universit\"at Stuttgart,
  Pfaffenwaldring 57, D-70550 Stuttgart, Germany} \author{M.~Weitz}
\address{Institut f\"ur Angewandte Physik der Universit\"at Bonn,
  Wegelerstra\ss e 8, D-53115 Bonn, Germany}

\date{\today}
  
\begin{abstract}
  We consider Feshbach scattering of fermions in a one-dimensional optical
  lattice. By formulating the scattering theory in the crystal momentum basis,
  one can exploit the lattice symmetry and factorize the scattering problem in
  terms of center-of-mass and relative momentum in the reduced Brillouin zone
  scheme.  Within a single band approximation, we can tune the position of a
  Feshbach resonance with the center-of-mass momentum due to the non-parabolic
  form of the energy band.
\end{abstract}
\maketitle



\section{Introduction}

It is well known from solid state theory \cite{slater58,callaway91}, x-ray
defraction \cite{batterman64}, or the atomic motion in laser fields
\cite{cohentannoudji92} that the presence of a periodic potential requires a
modification of the conventional concepts of scattering theory
\cite{taylor00}. In the context of ultra-cold quantum gases, this has led to a
tremendous outburst of activities, during the past years. Today, it is
possible to examine the interplay of many-body physics at the lowest
attainable temperature, in the presence of designable optical lattices
\cite{weitz04} for bosons \cite{jaksch98,kollath04,bloch05,oberthaler06},
fermions \cite{rigol03,rigol04,koehl05,orso05}, bose-fermi mixtures
\cite{carr05,ospelkaus06} and to manipulate simultaneously the interaction
among particles \cite{dickerscheid05,orso05b,stoeferle06,buechler06,wouters06},
as well as collective states \cite{grupp05}.

In the present article, we will discuss binary Feshbach resonance scattering
in the presence of a lattice, as a particular aspect of the aforementioned
general theme.  When one approaches the topic of binary atomic scattering in
homogeneous space and in a periodic lattice, one needs to highlight the
similarities and differences, first.  In homogeneous space the scattering
process connects asymptotical free states, which are plane waves $\ket{q}$.
The accessible relative kinetic energy $q^2>0$ forms a simple continuum, which
has a lower bound, but no upper one.  Furthermore, a binary scattering event
in homogeneous space is translational invariant.  As a consequence, one
obtains the separation of the center-of-mass motion from the relative
dynamics, hence a significant simplification of the problem.  In optical
lattices in contrast, we have to use the eigenstates of a lattice, \ie
Bloch-states $\ket{q,n}$ with a quasi-momentum $q$ and band-index $n$. Due to
the periodic potential, we find a structured energy continuum
$\epsilon_L(q,n)$, which consist of several bands $n$ of increasing width
(defining upper and lower band edges), as well as forbidden gaps. The density
of states also varies accordingly.

The discrete translational symmetry of the lattice can be exploited for the
scattering problem by introducing a crystal momentum basis \cite{adams52},
which is formed by two-particle free Bloch states $\ket{q_1, n_1}\otimes\ket{
  q_2, n_2}$. They are labeled by the center-of-mass momentum $Q=q_1+q_2$
modulo the crystal momentum $k_L$, which remains a good quantum number. In
this way, the coupled two-particle scattering simplifies again and we can
introduce a scattering amplitude to measure the strength of the binary
interaction.

By introducing the Feshbach coupling, in addition to the normal pairing
potential between two fermions, we will be able to enhance the strength of binary
interactions via an external magnetic field \cite{tiesinga92}. Being in a
lattice, we can consider interband interactions in addition to the intraband
transitions.  This will be important if the strength of the Feshbach coupling
is comparable to the interband separation. For simplicity, we will consider
here only very narrow resonances, such that a single band description is
sufficient. Extensions to multi-band configurations are straight forward in
the present formulation, if necessary \cite{diener06}.

In this article, we present the principle of a scattering calculation for two
fermions in a one-dimensional lattice. In Sec.~\ref{sec2}, we will start from
the current form of the many-body theory that is pursued by many groups. We
will simplify this by considering only two fermions and a compound bosonic
molecule to obtain a two-component two-particle Schr\"odinger equation for the
molecule and fermion-pair wave-function. In Sec.~\ref{sec3}, we will briefly
review the basic concept of the scattering phase, the transmission probability
and discuss the simplest model for a Feshbach resonance in homogeneous space.
This will be generalized to two-particle Feshbach scattering in a
one-dimensional lattice in Sec~\ref{sec4}. We present numerical results for
Feshbach resonance within a single band approximation and demonstrate that it
can be tuned selectively with the center-of-mass momentum, due to the
non-parabolic energy-band.

\section{Reducing many-body physics to two atoms and a molecule \label{sec2}} 
Currently, much effort is devoted to the many-body description of resonance
superfluidity and the BEC-BCS crossover \cite{regal03,chin04,zwierlein04}.
Thus, we will briefly introduce the fundamental model Hamiltonian
\cite{kokkelmans501,kokkelmans502}. Then, we will apply it to the situation of
only two fermions and a bosonic molecule, now, trapped in the same periodic
one-dimensional optical lattices. The periodic trapping of both molecular and
atomic components is beneficial to the overall interaction cross section as
both components will tend to be localized at the anti-nodes of the optical
potential, thus be constantly available for scattering.

In the language of second quantization, we describe the many-body system with
fermionic fields $\annhilate{\psi}{\sigma}(\boldvec x)$, which remove a single
fermionic particle from position ${\boldvec x}$ in internal state
$\sigma=\{\uparrow, \downarrow\}$, and molecular bosonic fields
$\annhilate{\phi}{i}(\boldvec x)$, which annihilate a composite bound
two-particle excitation from the center-of-mass space-point ${\boldvec x}$ in
internal configuration $i$.  These field operators and their adjoints satisfy
the usual fermionic anti-commutation rules
\begin{eqnarray}
  \left \{ \annhilate{\psi}{\sigma_1}({\boldvec x}_1),
    \create{\psi}{\sigma_2}({\boldvec x}_2) \right \}&=&
  \delta({\boldvec x}_1-{\boldvec x}_2)\,
  \delta _{\sigma_1 \sigma_2}, \nonumber\\
  \left \{\annhilate{\psi}{\sigma_1}({\boldvec x}_1),
    \annhilate{\psi}{\sigma_2}({\boldvec x}_2) \right \} &=&0,
\end{eqnarray}
and bosonic commutation rules
\begin{eqnarray}
\label{commuti}
\left [ \hat{\phi}_{i_1}^{\phantom\dag}({\boldvec x}_1),
    \hat{\phi}_{i_2}^{\dag}({\boldvec x}_2) \right ]&=&
  \delta({\boldvec x}_1-{\boldvec x}_2)\,
\delta_{i_1 i_2},\nonumber \\
\left[\hat{\phi}_{i_1}^{\phantom\dag}({\boldvec x}_1),
\hat{\phi}_{i_2}^{\phantom\dag}({\boldvec x}_2)\right]
 &=& 0,
\end{eqnarray}
respectively. We want to consider only bound molecular excitations by choosing
a high dissociation threshold energy. Effectively this closes this decay
channel and allows only for collision induced processes. These are the basic
ingredients of a Feshbach resonances as originally invented by E. Fermi,
U.~Fano and H. Feshbach \cite{Fano1961,Feshbach1962a,Feshbach1962b}. Hence, we
can present the composite molecular field also with respect to the individual
coordinates $(\bx_1,\bx_2)$ of the dimer, i.\thinspace{}e.
\begin{eqnarray}
  \annhilate{\phi}{}(\bx_1,\bx_2)=\sum_i
  \annhilate{\phi}{i}((\bx_1+\bx_2)/2) \langle \bx_2-\bx_1 |i \rangle
\end{eqnarray}

The dynamics of the multi-component gas is governed by a total system
Hamiltonian
\begin{eqnarray}
 \hat{H}=\hat{H}_L+\hat{V}\\
  \label{H0op}
  \hat{H}_L=\int \text{d}^3x \sum_{\sigma=\{\uparrow,\downarrow\}}
  \create{\psi}{\sigma}(\bx)
  H_L(\bx,\bp) \annhilate{\psi}{\sigma}(\bx) \nonumber\\
  +\int \text{d}^6x \,\create{\phi}{}(\bx_1,\bx_2)
  H_L(\bx_1,\bp_1,\bx_2,\bp_2) \annhilate{\phi}{}(\bx_1,\bx_2),\\
  \label{H1op}
  \hat{V}= \int \text{d}^6x\, \left\{
    v_{q}(\bx_1-\bx_2)\create{\phi}{}(\bx_1,\bx_2)
    \annhilate{\phi}{}(\bx_1,\bx_2) 
  \right. \nonumber\\
  + \left[  g^{\ast}(\bx_1-\bx_2)\create{\phi}{}(\bx_1,\bx_2)
    \annhilate{\psi}{\downarrow}(\bx_2)
    \annhilate{\psi}{\uparrow}(\bx_1)+{\rm H.c.}\right] \nonumber\\
  +\left.  v_{p}(\bx_1-\bx_2)
    \create{\psi}{\uparrow}(\bx_1)
    \create{\psi}{\downarrow }(\bx_2) 
    \annhilate{\psi}{\downarrow}(\bx_2) 
    \annhilate{\psi}{\uparrow }(\bx_1)  \right\}.
\end{eqnarray}
It consists of the lattice Hamiltonian $\hat{H}_L$ and the
interactions $\hat{V}$ between atoms and molecules. We assume that the free
dynamics of the atoms and molecules is determined by their kinetic and
potential energy in a quasi-1d optical lattice
\begin{eqnarray}
\label{freefermi}
H_L(\bx,\bp)=&-\frac{\hbar ^{2}}{2m}\nabla^{2}+U_L(x)+U_\perp(y,z),
\end{eqnarray}
where $\bx=(x,y,z)$ and $\bp=-i\hbar \nabla$ are canonically conjugate
variables in the position representation, $m$ is the atomic mass and
$U_{L}(x)$ is a one-dimensional optical lattice potential.  Furthermore, we
want to assume that the motion in the perpendicular $(y,z)$ direction is
effectively frozen out by a tight confinement potential $U_\perp(y,z)$.
Supposedly, the lattice energy is identical for the fermionic atoms of both
kinds and twice that for the molecules \ie
\begin{eqnarray}
\label{freemol}
H_L(\bx_1,\bp_1,\bx_2,\bp_2)=&H_L(\bx_1,\bp_1)+H_L(\bx_2,\bp_2).
\end{eqnarray}
The binary interaction potential $v_{p}$ accounts for the non-resonant
interaction of ``spin-up'' and ``spin-down'' fermions, the coupling strength
$g$ converts free fermionic particles into bound bosonic molecular excitations
and $v_q$ is the molecular potential with at least one bound state.  One can
further simplify matters by considering only even parity binary interactions
potentials $v_q$, $v_p$, and $g$.  Moreover, we have neglected the
interactions among the molecules, since we will focus on the case of just two
fermions.

The essence of the resonant scattering physics in this many-body Hamiltonian
can be brought out most clearly, if we consider only two interacting fermionic
atoms with field-amplitude $\psi$ and a bosonic molecule with amplitude
$\phi$, \ie
\begin{eqnarray}
  \ket{\Psi(t)}=& \int \text{d}^6x \left\{
    \phi(\bx_1,\bx_2,t) \create{\phi}{}(\bx_1,\bx_2)\right. \nonumber\\
  &\left.+\psi(\bx_1,\bx_2,t) \create{\psi}{\uparrow}(\bx_1)
    \create{\psi}{\downarrow }(\bx_2) \right\}\ket{0}.
\end{eqnarray}
According to the Pauli principle, the bosonic and fermionic part of the total
wave function must be symmetric and anti-symmetric under particle exchange.
Thus, if we limit the discussion to the fermionic singlet channel, we need to
have a symmetric spatial amplitude $\psi(\bx_1,\bx_2,t)= \psi(\bx_2,\bx_1,t)$,
as well as a symmetric molecular wavefunction
$\phi(\bx_2,\bx_1,t)=\phi(\bx_1,\bx_2,t)$.

In this restricted few-particle Fock space with the imposed constraints on the
interaction channels, finally one finds a two-component Schr\"odinger
equation for the state vector $\bchi(\bx_1,\bx_2,t)=
(\phi(\bx_1,\bx_2,t),\psi(\bx_1,\bx_2,t))^\top$
\begin{eqnarray}
  \label{chi100}
    i\hbar\partial_t \bchi=
    \left[H_L(\bx_1,\bp_1,\bx_2,\bp_2)\otimes\mathds{1}+
      V(\bx_1-\bx_2)\right]\bchi, \nonumber\\
    V(\bx)=\left(\begin{array}{cc}
      v_q(\bx) &   g(\bx)  \\
      g^{\ast}(\bx) & v_p(\bx)
    \end{array}\right).
\end{eqnarray}
In the following, we will also abandon the three-dimensional character of the
problem and restrict the discussion to a quasi one-dimensional situation, when
all dynamics takes place along the $x$-direction.

\section{Scattering in free space\label{sec3}}
The previously introduced two-component Hamiltonian contains the essential
ingredients of Feshbach scattering, but lacks the full translational
invariance of homogeneous space, due to the presence of the lattice potential.
We will therfore briefly review the scattering phase and the prerequisites for
the appearance of a Feshbach resonance in free space with a contact
potential for later reference.
\subsection{Scalar potential scattering}
The energy of two particles on a line
\begin{eqnarray}
  \label{ham0}
  H=H_0(p_1)+H_0(p_2)+V(x_1-x_2),
\end{eqnarray}
consists of kinetic energy $H_0=p^2/2m$ and the short-range binary potential
energy $V$. Obviously, all parts of this Hamiltonian are translational
invariant.  This symmetry can be exploited by introducing a center-of mass
coordinate $X$ and a relative coordinate $x$, as well as their conjugate
momenta $P,p$ as
 \begin{eqnarray}
X&=&(x_1+x_2)/2,  \quad x=x_1-x_2,\\
P&=&p_1+p_2,    \quad p=(p_1-p_2)/2.
\end{eqnarray}
The total momentum $P$ is the conserved quantity that is associated with the
symmetry generating displacement operator $\mathcal{D}_{12}(a)=\exp{[i a
  P/\hbar]}$. It shifts the whole system by a distance $a$ in the
$\mathbf{e}_{12}= \mathbf{e}_1+\mathbf{e}_2$ direction and commutes with the
Hamiltonian
\begin{eqnarray}
  [H, \mathcal{D}_{12}(a)]=0.
\end{eqnarray}
Thus, the total two-body wave-function $\chi_2(x_1,x_2)$ can be expressed in
terms of center-of-mass and relative coordinates with a product ansatz
$\chi_2(x_1,x_2)=\theta(X)\chi(x)$. Here, we assume that $\theta(X)=\exp{(iQ
  X)}$ is a plane wave in the center-of-mass coordinate with momentum $Q$ and
the relative wave-function is denoted by $\chi(x)$. This reduces the
corresponding two-particle Schr\"odinger equation with total energy $E$ to the
standard form \cite{taylor00} of an effective single particle problem
\begin{equation}\label{schroed0}
\left[\frac{d^2}{dx^2}+\varepsilon-V(x)\right]\chi(x)=0,
\end{equation}
where we have also rescaled all dimensional quantities, like length
$x\rightarrow x k_L$, in terms of the wave-number $k_L$ of an optical lattice
photon [see \eq{1820}], as well as energies or
potentials $V\rightarrow V/\varepsilon_L$, in terms of the recoil energy
$\varepsilon_{L}=\hbar^2 k_L^2/m$. Adopting such units, will facilitate the
comparison with the lattice case discussed in Sec.~\ref{sec4}. Then, the relative energy
$\varepsilon$ becomes synonymous with the free space dispersion relation
\begin{eqnarray}
\label{releng}
\varepsilon(k)=k^2=E-\frac{Q^2}{4}>0.  
\end{eqnarray}

As any second order differential equation, \eq{schroed0} admits two linearly
independent scattering solutions $\chi^{(\pm)}_k(x)$ for the same energy
$\varepsilon(k)$. By comparing them to non-interacting plane waves, one can
introduce reflection and transmission amplitudes $R(k)$ and $T(k)$, \ie
\begin{eqnarray}
    \lim_{x \rightarrow -\infty}
    \chi^{(+)}(x,k)&=&e^{ikx}+R\,e^{-ikx} \nonumber\\
    &=&e^{i\delta_e}\cos(kx-\delta_e)+ie^{i\delta_o}\sin(kx-\delta_o),\\
    \lim_{x\rightarrow\infty}\chi^{(+)}(x,k)&=&T\,e^{ikx} \nonumber\\
    &=&e^{i\delta_e}\cos(kx+\delta_e)+ie^{i\delta_o}\sin(kx+\delta_o) 
\end{eqnarray}
as well as even and odd scattering phase shifts $\delta_e(k)$ and
$\delta_o(k)$. The transmission and reflection probabilities are given by
\begin{equation}
  \label{T1}
    |T|^2=\cos^2(\delta_e-\delta_o),\quad
    |R|^2=\sin^2(\delta_e-\delta_o),
\end{equation}
and unitarity demands current conservation, which is mathematically paraphrased
as $|T|^2+|R|^2=1$.

\begin{figure}[h]
  \centering
  \includegraphics[angle=-90,width=0.7\columnwidth]{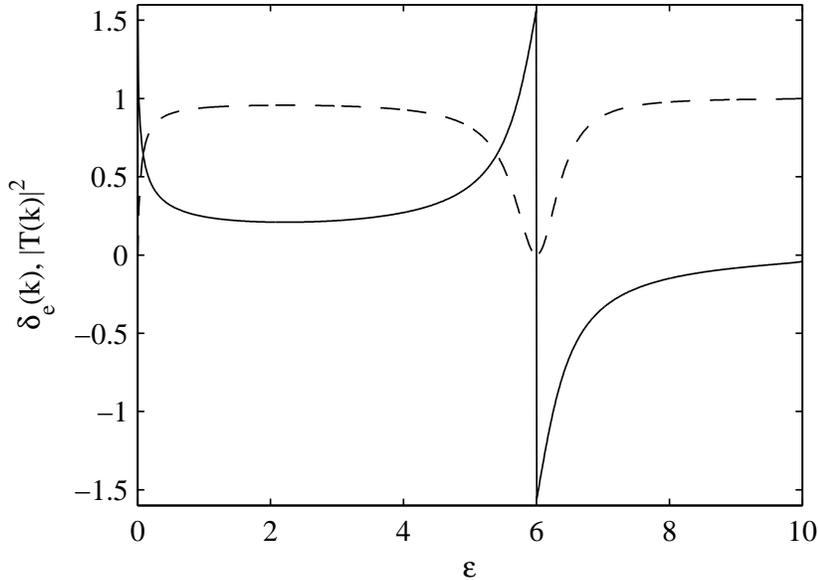}
  \caption{Even scattering phase $\delta_e(k)$ (solid line) and 
    transmission probability $|T|^2$ (dashed line) of a Feshbach resonance as
    a function of energy $\varepsilon=k^2$ with a resonance at
    $\varepsilon_{res}=6$.  Parameters: $u=10$, $v=-4$ and $g=1$.
    \label{1}}
\end{figure}
\subsection{Two-component Feshbach resonances with a contact potential}
The quintessential mechanism for a Feshbach resonance occurs, when one couples
two internal states $\bchi=(\chi_q,\chi_p)^\top$ to the external motion of the
one-dimensional, two-particle Hamiltonian of \eq{ham0}.  The closed q-channel
needs an attractive delta potential to support a bound state and a localized
coupling matrix element is required to provide an interaction with the open
p-channel \cite{tobocman61}. A simple, analytically solvable model for this is
\begin{eqnarray}
  \label{r11}
    H=H_0(p_1,p_2)\otimes\mathds{1}+V(x_1-x_2), \nonumber\\
    V(x)=\left(\begin{array}{cc}
      u+v \delta(x)& g \delta(x) \\
      g \delta(x) & 0
    \end{array}\right).
\end{eqnarray}
The parameters of the potential matrix $V$ are the threshold energy $u>0$, the
strength of the closed channel potential $v<0$ and the channel coupling
parameter $g$.  As in the scalar case, all contributions to the Hamiltonian
are translationally invariant and a product ansatz for the wave-function
$\bchi_2(x_1,x_2)=\exp{(iQ X)}\bchi(x)$ leads to the Schr\"odinger equation
for the relative wave-function
\begin{equation}
  \label{r1}
  \left[\frac{d^2}{dx^2}+\varepsilon-
    V(x)\right]
{\bchi}(x)=0. 
\end{equation}
Here, we have again introduced rescaled potentials, energies and length
mentioned in the context of \eq{schroed0}.

For relative scattering energies $0<\varepsilon(k)=k^2<u$, smaller than the
threshold energy $u$, the closed channel asymptotic wave-function $\chi_q(x)$
vanishes exponentially, while the open channel wave-function $\chi_p(x)$
propagates outward freely. Due to the even parity of the contact potential, we
can also choose wave-functions of definite parity, \ie
\begin{eqnarray}
    \lim_{x\rightarrow\pm\infty}\chi_q^{(e)}(x,k)\sim e^{- q |x|},\quad
    q(k)=\sqrt{u-k^2},\\
    \lim_{x\rightarrow\pm\infty}\chi_p^{(e)}(x,k)\sim \cos(kx\pm\delta_e).
\end{eqnarray}
Odd wavefunctions do not accrue any phase shift, as they vanish at the point
of interaction.
\begin{figure}[h]
  \centering
  \includegraphics[angle=-90,width=0.7\columnwidth]{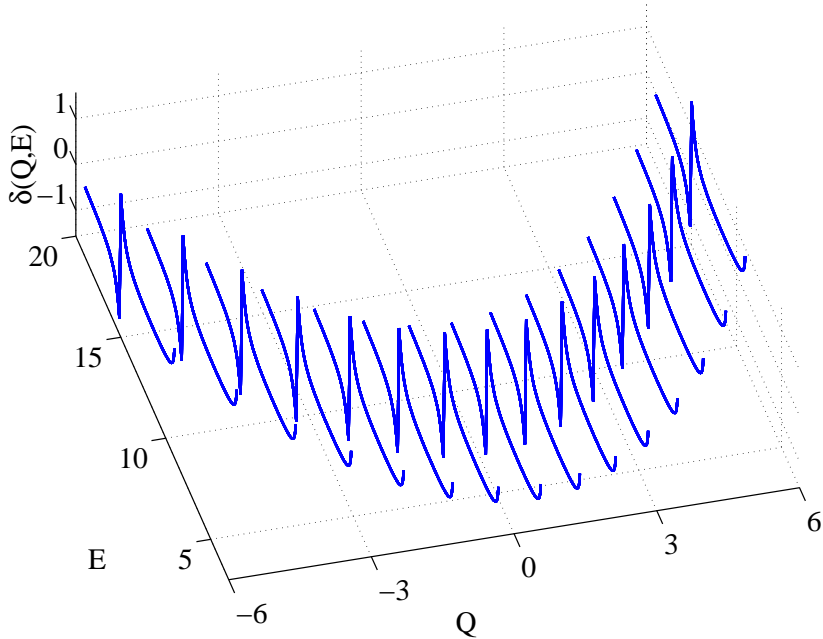}
  \caption{Even scattering phase $\delta_e(k(Q,E))$ 
    versus energy $E$ and momentum $Q$. The Feshbach resonances are located on
    a free space parabola as a function of $Q$.  Parameters as in \eq{param}.}
  \label{Kmathom}
\end{figure}

By integrating the coupled Schr\"odinger equation \eq{r1} over an
infinitesimal strip around the discontinuity, one obtains the necessary
boundary conditions to compute the scattering phase of the even wavefunction.
After some minor algebra, one obtains the phase shifts
\begin{eqnarray}
  \label{feshbach}
  \tan{(\delta_e)}&=-\frac{g^2}{2k(2q(k)+v)},\qquad \delta_o=0 .
\end{eqnarray}
This phase shift and the corresponding transmission probability $|T|^2$ of
\eq{T1} are depicted in Fig.~\ref{1}. One clearly observes the Feshbach
resonance with a $\pi$-phase jump at $\varepsilon_{res}=k^2_{res}=u-v^2/4$.
The position of the resonance energy is usually controlled by changing the
dissociation threshold energy $u$ via magnetic fields or via the depth of the
bound state $v$, which is an property of the considered atomic element and,
thus,  harder to modify. The width of the resonance is primarily determined by
the coupling strength $g$ \cite{friedrich}.  Due to the vanishing odd
phase-shift, the total reflection $T=0$ also occurs at the same energy, when
the even phase shift reaches $\delta_e(k_{res})=\pi/2$.  It is also important
to note that the phase can be well approximated by a Breit-Wigner curve in the
vicinity of the resonance.

In Fig.~\ref{Kmathom}, we present the same scattering phase $\delta_e(k(E,Q))$
of \eq{feshbach}, now as a function of the two-particle energy $E$ and the
center-of-mass momentum $Q$, according to \eq{releng}. This representation
emphasizes the parabolic form of the resonance energy
$E_{res}=\varepsilon_{res}+Q^2/4>0$, as a function of $Q$.  The Feshbach
resonance is always present for $\varepsilon_{res}>0$ and for all $Q$, or it
disappears if $\varepsilon_{res}<0$. As we will show in the following sections,
this fundamental behavior can be changed by considering Feshbach scattering
in a lattice where center-of-mass and relative motion are coupled. Results can
be seen in the final Figs.~\ref{Kmat} and \ref{Kmat2}.
\section{Scattering on a one-dimensional lattice\label{sec4}}
\subsection{Bloch states  and periodic boundary conditions}
Studying physics in periodic structures immediately leads to the consideration
of Bloch states \cite{kohn59,callaway91}, which reflect the discrete
translation symmetry $\mathcal{D}(l a)=\exp{[ilap/\hbar]}$ of the Hamiltonian
$[H_L,\mathcal{D}(l a)]=0$ with $l\in \mathds{N}$ .  We have already
introduced a quasi-one dimensional lattice Hamiltonian in \eq{freefermi} and
will disregard from now on the transverse degrees motion, \ie
\begin{eqnarray}
  \label{1820}
  H_L(x,p)&=&\frac{p^2}{2m}+U_L(x), \quad U_{L}(x)=U_L\cos^2{(k_L x/2)}.
\end{eqnarray} 
The lattice potential $U_L(x+a)=U_L(x)$ is characterized by a lattice
constant $a$, which in turn defines the crystal momentum $k_L=2\pi/a$.
Thus, the eigenstates of Schr\"odinger equations
\begin{equation}\label{182098}
  H_L|q,n\rangle=\varepsilon_L^n(q)|q,n\rangle,
\end{equation}
can be classified as Bloch states $\ket{{q,n}}$ with quasi momentum
$-\pi/a\leq q< \pi/a$, and energy bands $\varepsilon_L^n(q)$, labeled by
band-index $n$. According to the Bloch theorem 
\begin{equation}\label{15}
\scp{x}{{q,n}}=e^{iqx}u_{q}^{n}(x),\quad  u_q^n(x+a)=u_q^n(x).
\end{equation}
Such an energy eigenfunction can be decomposed into a plane wave phase factor
and a lattice periodic function $u_q^n$.  Subjected to translation to the next
lattice site, the wave-function acquires a complex phase
\begin{equation}
  \label{1811}
    \scp{x+a}{{q,n}}=e^{iqa}\scp{x}{{q,n}}.
\end{equation} 
The periodic continuation in momentum space has its subtleties and attention
needs to be paid to the vanishing of the wave-function at the center or edge
of the first Brillouin zone \cite{kohn59}.  However, the generic case is
simply given by $\scp{x}{{q+k_L,n}}=\scp{x}{{q,n}}$.
\begin{figure}[h]
  \centering
  \includegraphics[angle=-90,width=0.7\columnwidth]{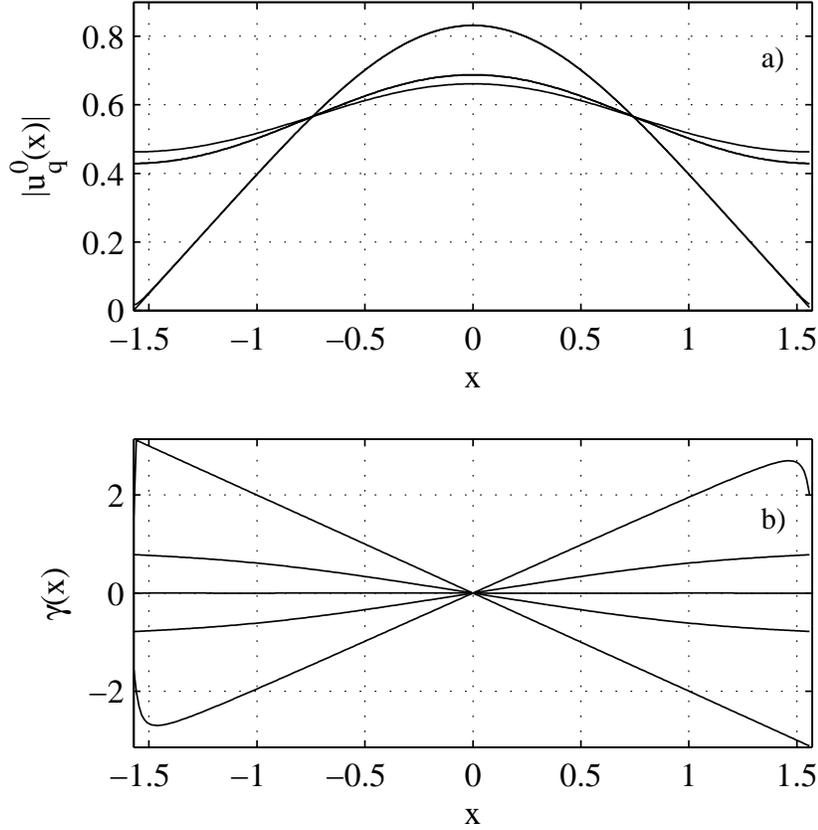}
  \caption{Magnitude $|u_q^0(x)|$ (a) and phase $\gamma(x)$ 
    (b) of the lowest band Bloch function $\scp{x}{q,n=0}=|u_q^{n}(x)|
    \exp{[i\gamma (x)]}$, versus dimensionless position $-\pi/2\leq x<\pi/2$,
    for a few momenta $q=(-1, -0.5, 0, 0.5,0.992)$. In this shallow potential
    the Bloch waves are almost plane and deviations are only seen at the band
    edge $q\approx \pm 1$. Parameters as in \eq{param}.}
  \label{fig:0}
\end{figure}

In practice, it is usually necessary to work with a discrete subset of Bloch
states, which are found by considering a periodically continued, finite
lattice with an even number of wells $N=2 M$. The Born-von-Karman periodic
boundary conditions, then require that
\begin{equation}
  \label{181121}
  \scp{x+L}{{q,n}}=e^{iqL}\scp{x}{{q,n}}=\scp{x}{q,n}.
\end{equation} 
where the total length $L=Na$. This can only be true, if there are
exactly $N$ distinct momenta $q_l$ in each band
\begin{equation}\label{3812}
- \frac{\pi}{a}\le q_l=\frac{2\pi l}{L}=k_L\frac{ l}{N}<\frac{\pi}{a}
,\quad -\frac{N}{2}\le l <\frac{N}{2}.
\end{equation}
We choose the following orthogonalization for the Bloch-states on a finite
lattice
\begin{eqnarray}
      \scp{q_1,n_1}{q_2,n_2}=N \delta_{q_1,q_2} \delta_{n_1,n_2}, \quad\int_{0}^{a}dx |\scp{x}{q,n}|^2=1.  
\end{eqnarray}

\begin{figure}[h]
  \centering 
  \includegraphics[angle=-90,width=0.6\columnwidth]{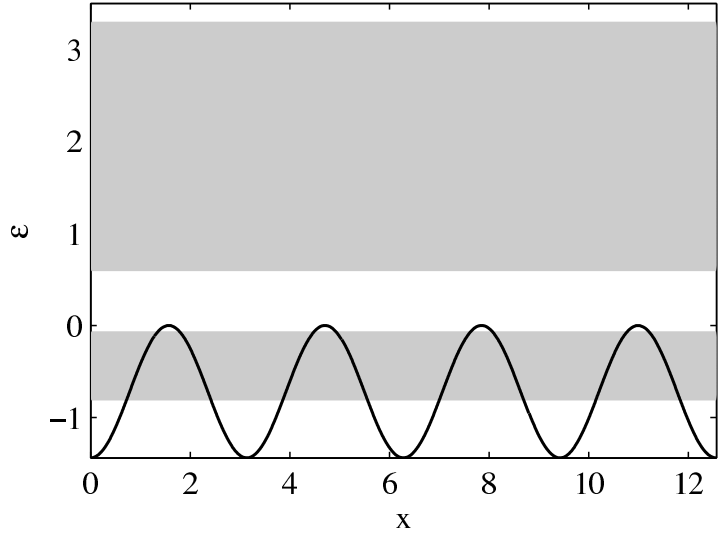}
  \caption{Lattice potential $U_{L}(x)$  (solid line) versus  position 
    $x$, superimposed on top of two allowed energy bands
    $\varepsilon_L^0(q)$ and $\varepsilon_L^1(q)$ (shaded gray).
    Parameters as in \eq{param}.}
  \label{fig:2a}
\end{figure}

\begin{figure}[h]
  \centering 
  \includegraphics[angle=-90,width=0.6\columnwidth]{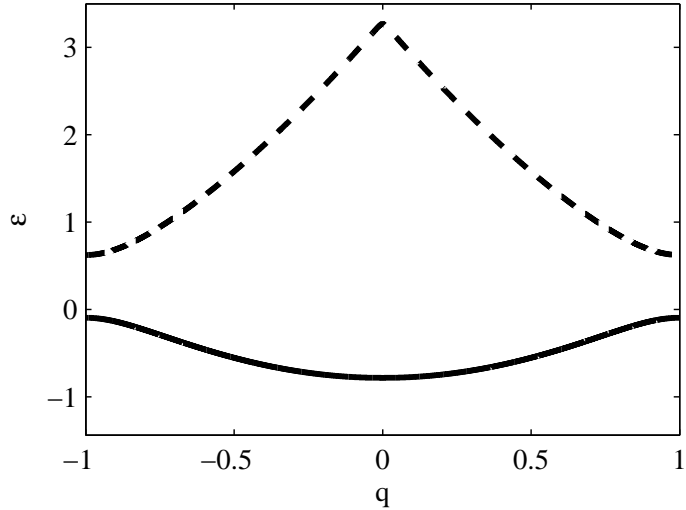}
  \caption{Energy bands $\varepsilon_L^0(q)$ (solid) and 
    $\varepsilon_L^1(q)$ (dashed) versus quasi-momentum $q$ in the first
    Brillouin zone $-1\leq q< 1$.}
  \label{fig:2b}
\end{figure}

The general behavior of the Bloch states is shown in Fig.~\ref{fig:0}. There
we have selected a few Bloch eigenfunctions for some momenta $q$ within the
Brillouin zone, which exemplify the modified plane wave behavior.  A typical
energy-band structure in coordinate and momentum space is presented in
Fig.~\ref{fig:2a} and Fig.~\ref{fig:2b}, respectively.  In here and all of the
following calculations, we use dimensionless parameters for a shallow lattice
that only supports one band below the barrier \ie
\begin{eqnarray}
\label{param}
  U_L=-(6/5)^2, \quad a=\pi, \quad N=2^8.
\end{eqnarray}
\subsection{Scalar two-particle scattering in lattices}
We will now consider the scattering of two particles in the presence of a
lattice. Thus, we have a two-particle lattice Hamiltonian and a binary
interaction $V$
\begin{eqnarray}
  \label{1800}
    H=H_L(x_1,p_1)+H_L(x_2,p_2)+V(x_1-x_2).
\end{eqnarray} 
By forming two-particle basis states out of single particle Bloch waves one
obtains two-dimensional Bloch states
\begin{eqnarray}
  \label{191433}
  |Q,q,n_1,n_2\rangle=\ket{q_1(Q,q),n_1}  \otimes\ket{q_2(Q,q),n_2}\\
    \label{191}
    Q=q_1+q_2, \quad q=(q_1-q_2)/2\\
    q_1=Q/2+q, \quad q_2 =Q/2-q.   
\end{eqnarray}
\begin{figure}[h]
  \centering \includegraphics[angle=0,width=0.7\columnwidth]{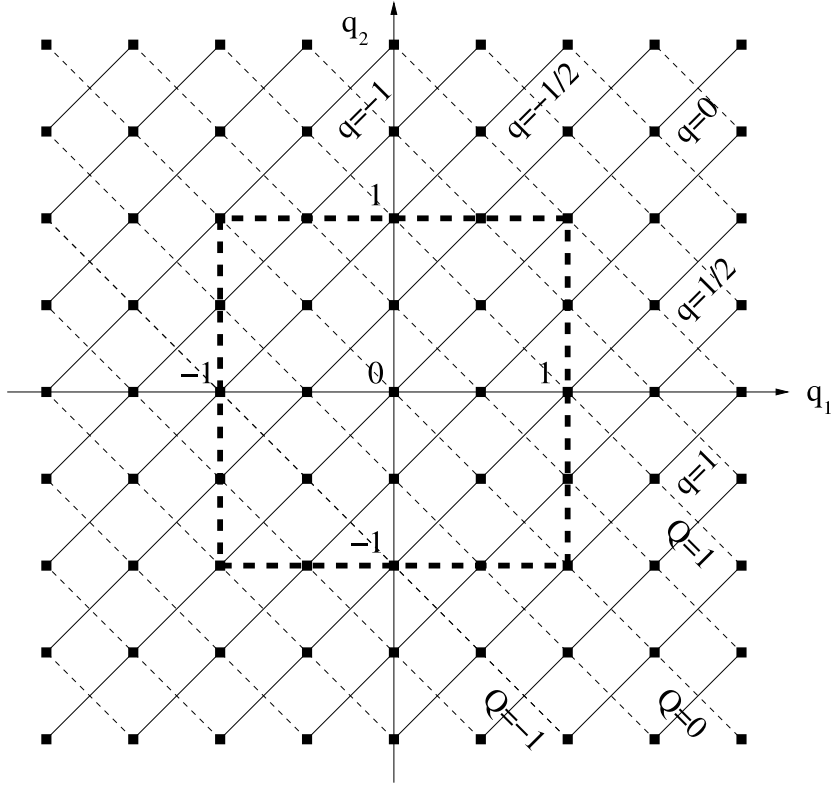}
  \caption{Reciprocal momentum space for two particles in a one-dimensional lattice consisting
    of only $N=4$ wells. The Cartesian axes represent the individual momenta
    in the reduced zone scheme $-1\le q_1,q_2<1$. The $45^\circ$ rotated grid
    lines correspond to the contours of center-of-mass $-1\le Q<1$ and
    relative momentum $-1 \leq q <1$.}
  \label{fig:3}
\end{figure}

The corresponding reciprocal momentum space for two particles in a one-dimensional lattice 
is depicted in Fig.~\ref{fig:3}.  One can see the individual single particle momenta $q_1$
and $q_2$, as well as the center-of-mass momentum $Q$ and relative momentum
$q$. This set of symmetry adjusted basis states can be used to represent a
general quantum state with quasi-momentum $Q$ as
\begin{eqnarray}
  \label{23}
  |\chi,Q\rangle=\sum_{q,n_1,n_2} \chi_{Q}^{(n_1,n_2)}(q)|Q,q,n_1,n_2\rangle\\
  \label{22}
  \mathcal{D}_{12}(la)|\chi,Q\rangle=e^{ila Q}|\chi,Q\rangle.
\end{eqnarray} 
From the discrete translation symmetry of the system, one finds the selection
rule
\begin{eqnarray}
    0&=\bra{\chi,Q}[H,\mathcal{D}_{12}(la)]\ket{\chi,Q^\prime}=
\bra{\chi,Q}H\ket{\chi,Q^\prime}(e^{ila(Q^\prime-Q)}-1),
\end{eqnarray}
which implies $Q^\prime=Q \text{mod} k_L$.  In the following, we will consider
only transitions within the lowest energy band \ie $n_1=n_2=0$ and simplify
the notation to $\chi_{Q}^{(0,0)}(q)\equiv \chi_{Q}(q)$, consequently.
\begin{figure}[h]
  \centering 
  \includegraphics[angle=0,width=0.7\columnwidth]{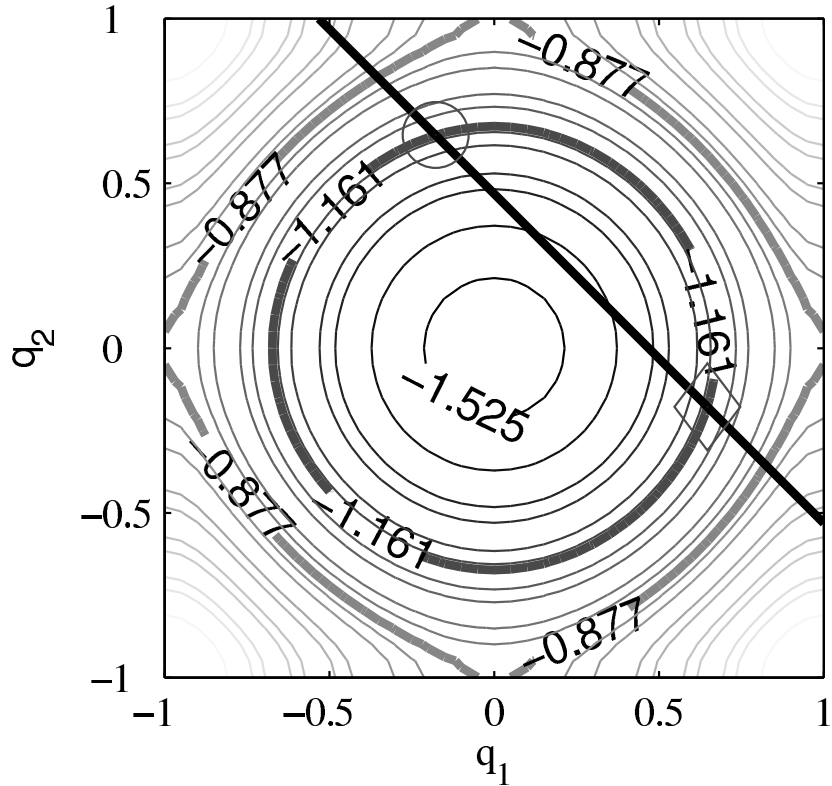}
  \caption{Energy contours $E_L(Q,q)$ for the
    non-interacting two-particle system in momentum space. The superimposed
    $Q=$const. line (solid black) has two intersections with another
    $E=$const. contour (dark gray), which are at $q$ (circled) and $-q$
    (boxed).}
  \label{figcontour}
\end{figure}

\begin{figure}[h]
  \centering \includegraphics[angle=-90,width=0.7\columnwidth]{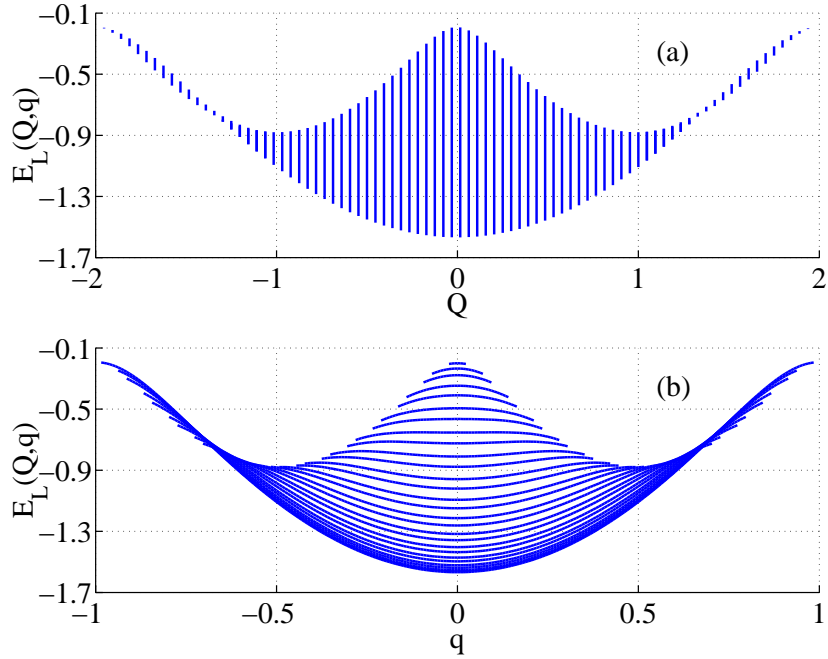}
  \caption{Projections of the two-dimensional energy surface $E_L(Q,q)$ 
    versus momentum $Q$ (a) and $q$ (b).}
  \label{fige0}
\end{figure}

Within these assumptions, one can represent the Schr\"odinger equation to
\eq{1800} as
\begin{eqnarray}
  \label{285}
  0=&\sum_{q'}\Big\{[\varepsilon_L^0(q_1)+
  \varepsilon_L^0(q_2)-E]\delta_{qq^\prime}
  +V_Q(q,q')\Big\}\chi_{Q}(q'),
\end{eqnarray}
where $q_1=q_1(Q,q)$ and $q_2=q_2(Q,q)$.  If we let our perturbation tend to
zero, then we are left with the two-dimensional energy surface
\begin{equation}
  \label{2852}
  E_L(Q,q)=\varepsilon_L^0(q_1)+\varepsilon_L^0(q_2).
\end{equation}
The contour lines of which are depicted in Fig.~\ref{figcontour}. The marked
intersections show that there are two linearly independent two-particle
quantum states labeled by $(q_1,q_2)$ and $(q_2,q_1)$ that have the same total
energy $E$ and center-of-mass momentum $Q$.  It it interesting to note that in
general there are no straight contour-lines at $Q=\pm1$ or $q=\pm1$, as would
be the case in the tight binding limit $|U_L| \gg 1$. The energy range that is
covered by the two-particle energy can be seen in Fig.~\ref{fige0}. It shows
the projections along the $Q$ and $q$ momentum lines.

\begin{figure}[h]
  \centering
  \includegraphics[angle=-90,width=0.7\columnwidth]{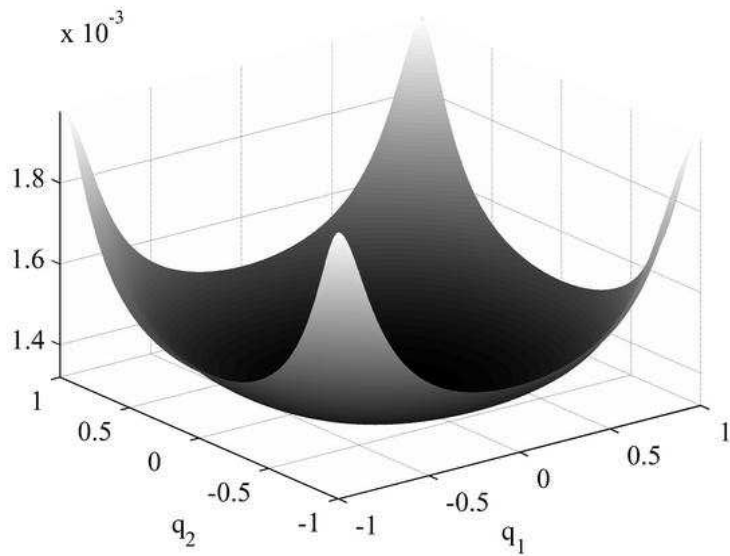}
  \caption{Interaction matrix element $v_Q(q,q')$ in reciprocal momentum space
    for $Q=0$, other parameters as in \eq{param}.} \label{fig:4a}
\end{figure}

Now, if we turn to the evaluation of the matrix element $V_Q(q,q')$, then it is
best expressed in center-of-mass coordinates. Moreover, the algebra simplifies
considerable, if we use a zero-range contact potential for the binary
potential $V(y)=v\delta(y)$ and obtain
\begin{eqnarray}
  \label{2834}
  V(Q,q,q')&=&v v_Q(q,q'),\\
  v_Q(q,q')&=&
  \int_0^a \frac{dx}{N} \Big\{u^*_{Q/2+q}(x) u^*_{Q/2-q}(x) 
  u_{Q/2+q'}(x) u_{Q/2-q'}(x) \Big\}.
\end{eqnarray}
The shape of this potential matrix can be calculated quite accurately with
localized Wannier functions in the Gaussian approximations. In general, this
agrees well with the numerical results shown in Fig.~\ref{fig:4a}. In this
picture, we have chosen a value of $Q=0$ for the center-of-mass momentum. For
other values $Q\neq0$, one obtains a modest variation of the shape of the
matrix element at the momentum edges, but it remains predominantly constant
inside.

\subsection{Two component Feshbach scattering in lattices}
Having established the basic notions and concepts in the previous sections, we
can now turn to the Feshbach resonance scattering phenomenon in a lattice. The
basic Hamiltonians have been already introduced in Eqs.~(\ref{chi100}),
(\ref{r11}) and (\ref{1800}).  Explicitly, we have the motion in the lattice
Hamiltonian $H_L$, as well as the Feshbach potential $V$
\begin{eqnarray}
  \label{27222}
  H=H_L(x_1,p_1,x_2,p_2)\otimes \mathds{1}+V(x_1-x_2)\nonumber\\
  V(x)=\left(\begin{array}{cc}
    u+v \delta(x)& g \delta(x) \\
    g \delta(x) & 0
  \end{array}\right).
\end{eqnarray}
Here we have deliberately disregarded the pairing potential $v_p=0$, as it
only provides a modification to the Feshbach phenomenon.  The state of the
coupled two-component system in the lowest Bloch band $n=0$ is now defined as
\begin{eqnarray}
  \label{27223}
    |\bchi,Q\rangle
    =&\sum_{q} 
    \left(\begin{array}{cc}
        \phi_Q(q)\\
        \psi_Q(q)
      \end{array}\right)
    |Q,q\rangle,
\end{eqnarray}
where we have characterized the state with the momentum $Q$ and labeled the
open and closed channels by the molecular $\phi$ and two-fermion wave-function
$\psi$ as in \eq{chi100}.

In order to obtain the Bloch representation of the 
Schr\"odinger equation to \eq{27222} 
\begin{eqnarray}
  E\ket{\boldsymbol{\chi},Q}=H\ket{\boldsymbol{\chi},Q},
\end{eqnarray}
we project it on the two-particle Bloch eigenstate and get
\begin{eqnarray}
  E\left(\begin{array}{cc}
    \phi_Q(q)\\
    \psi_Q(q)
  \end{array}\right)=
  \sum_{q'}\Big\{
  [\varepsilon_L^0(q_1(Q,q))+\varepsilon_L^0(q_2(Q,q))]\delta_{q,q'}\nonumber\\
  \left.+
    \delta_{qq'}\left(
    \begin{array}{cc}
      u& 0 \\
      0 & 0
    \end{array}\right)+
    v_Q(q,q') 
    \left(\begin{array}{cc}
      v& g \\
      g& 0
    \end{array}\right)
  \right\}
  \left(\begin{array}{c}
    \phi_Q(q')\\
    \psi_Q(q')
  \end{array}\right).
\end{eqnarray}

\begin{figure}[h]
  \centering \includegraphics[angle=-90,width=0.7\columnwidth]{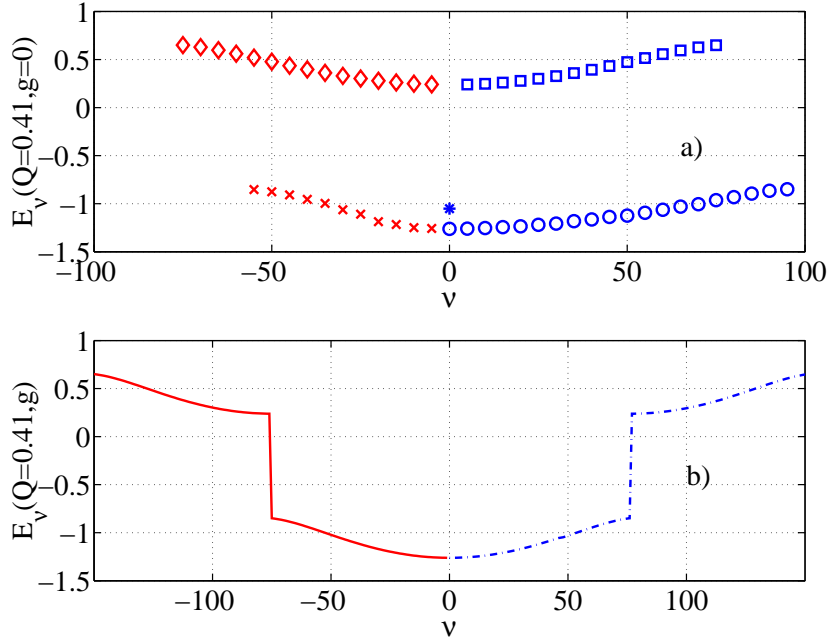}
  \caption{Discrete eigenenergies $E_\nu(Q,g)$ versus quantum number $-N\le
    \nu<N$ for momentum $Q=0.41$. Negative quantum numbers label
    odd states, while positive numbers count even states. a) Energies of
    uncoupled system $E_\nu(Q,g=0)$. Note the location of the bound state
    ($\ast$) of the molecular manifold ($\Diamond$, $\square$) is inside the
    fermionic spectrum marked by ($\times$, $\circ$). b) All interacting
    eigenenergies $E_\nu(Q,g=0.03)$ versus quantum index $\nu$.  Odd states
    (solid) and even states (dashed dotted). The bound state is now an
    admixture of molecular and fermionic manifolds.}
  \label{eng2comp}
\end{figure}

We have numerically diagonalized this Schr\"odinger equation for the lattice
parameters of \eq{param} and the Feshbach parameters $u=1.5$, $v=-7$,
$g=0.03$. Due to the parity of the Hamiltonian, one can sort the resulting $2
N$ eigenstates into even and odd states and assign them a positive or negative
quantum index $-N\leq \nu <N$.  By turning off the coupling between the
manifolds \ie $g=0$, one can distinguish easily the eigenvalues $E_\nu(Q,g=0)$
that belong to molecular (q-channel) or the open fermionic spectrum
(p-channel). We have depicted a few representative values in
Fig.~\ref{eng2comp}a). It can be seen clearly that there is a single bound
state embedded in the allowed energy band of the open p-channel.  In
Fig.~\ref{eng2comp}b), we present the interacting eigenvalues
$E_\nu(Q,g=0.03)$, which are now formed by admixtures of both manifolds and
can only be distinguished by the parity of the state.

In order to describe the scattering resonance, we use the standing wave version
of the Lippman-Schwinger equation \cite{taylor00}
\begin{eqnarray}
  \ket{\boldsymbol{\chi}^{(s)},Q,q}=
  \mathbf{e}_p\otimes \ket{Q,q}+G^{(s)}_L(E)V
  \ket{\boldsymbol{\chi}^{(s)},Q,q},
\end{eqnarray}
where $\mathbf{e}_p \otimes \ket{Q,q}$ represents a non-interacting Bloch-wave
in the asymptotically open p-channel $\mathbf{e}_p=(0,1)^\top$. For the
lattice Hamiltonian $H_L$, one obtains a standing wave Greens function
$G^{(s)}_L$ from the usual retarded and advanced Greens functions
$G^{(\pm)}_L$ by
\begin{eqnarray}
  G^{(s)}_L(E)&=\frac{1}{2}(G^{(+)}_L(E)+G^{(-)}_L(E)),\\
  G^{(\pm)}_L(E)&=\lim_{\epsilon\rightarrow 0+}\frac{1}{E \pm i\epsilon-H_L}.
\end{eqnarray}
Finally, the effect of scattering is measured by the off-shell Heitler-
or K-matrix element 
\begin{eqnarray}
  K_Q(q',q)= \bra{Q,q'}\mathbf{e}_p^\top V \ket{\boldsymbol{\chi}^{(s)},Q,q}.
\end{eqnarray}
We assign an even on-energy shell scattering phase shift
\begin{eqnarray}
    \tan{[\delta_e(Q,E)]}=K_Q(q,q'),\quad
    E=E(Q,q)=E(Q,q').     
\end{eqnarray}

\begin{figure}[ht]
  \centering \includegraphics[angle=-90,width=0.7\columnwidth]{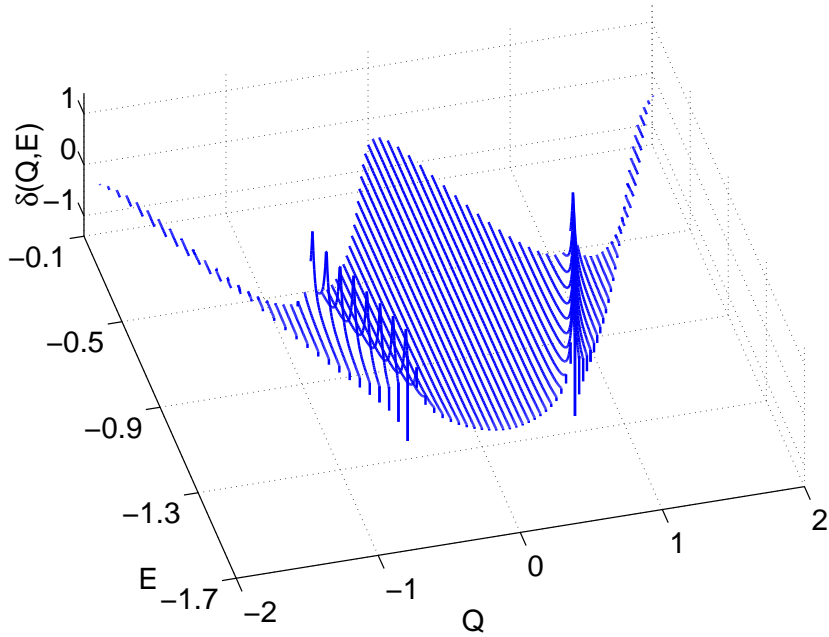}
  \caption{Even scattering phase-shift $\delta_e(E,Q)$ of a two-particle
    Bloch-wave versus energy $E$ and momentum $Q$. Once can see clearly
    Feshbach resonances symmetrically located around the $Q$-axis, which
    disappears when the bound state is outside the band. Parameters: $u=1.5$,
    $v=-7$, $g=0.03$.}
  \label{Kmat}
\end{figure}

We have evaluated this phase shift for two differently strong bound molecular
states and varied the center-of-mass momentum of the two-particle collision.
In the first Fig.~\ref{Kmat}, one can see the Feshbach resonance cutting into
the energy band and disappearing once it is outside.  This is modified in
Fig.~\ref{Kmat2}, which has a lesser bound resonance, and therfore does not
leave at the lower band edge.

This has to be compared with the Feshbach scattering in homogeneous space
according to \eq{feshbach} and has been presented already in
Fig.~\ref{Kmathom}. There, the resonance energy $E=\varepsilon_{res}+Q^2/4>0$,
is on a simple parabola as a function of $Q$. If we tune the Feshbach
resonance energy $\varepsilon_{res}<0$, it either disappears completely or is
always present.

\section{Experimental considerations}
Starting from a dilute Fermi gas, atoms can be loaded adiabatically into the
lowest energy band of the lattice. For quasi-momenta not too close to the band
edge, this requires that the Fermi energy remains clearly below the atomic
recoil energy $\hbar \omega _{r}$, corresponding to 2$\pi \cdot 8.4$~kHz and
2$\pi \cdot 74$ kHz for the $^{40}$K and $^{6}$Li D2-lines respectively
\footnote{Here $\hbar \omega_{r}=\hbar^2 k^{2}/2m \equiv \varepsilon_r/2$
  represents half the energy unit defined \eq{schroed0}.}. To ensure that the
Feshbach resonance provides coupling only within a single energy band, its
width has to be sufficiently narrow, \ie  below a value of about $\hbar
\omega _{r}/\Delta \mu ,$ where $\Delta \mu $ denotes the difference of the
atomic and the molecular states magnetic moments respectively. For typical
parameters, we arrive at a required width of the Feshbach resonance below 10
to 100 mG. We are aware, that the investigation of such weak Feshbach
resonances requires a high magnetic field stability.  On the other hand,
three-body losses, which for free space experiments are a mayor second
experimental constraint in the study of narrow Feshbach resonances, are
expected to be reduced in the presence of the lattice potential due to atom
localization in the micropotentials.

\begin{figure}[h]
\centering
\includegraphics[angle=-90,width=0.7\columnwidth]{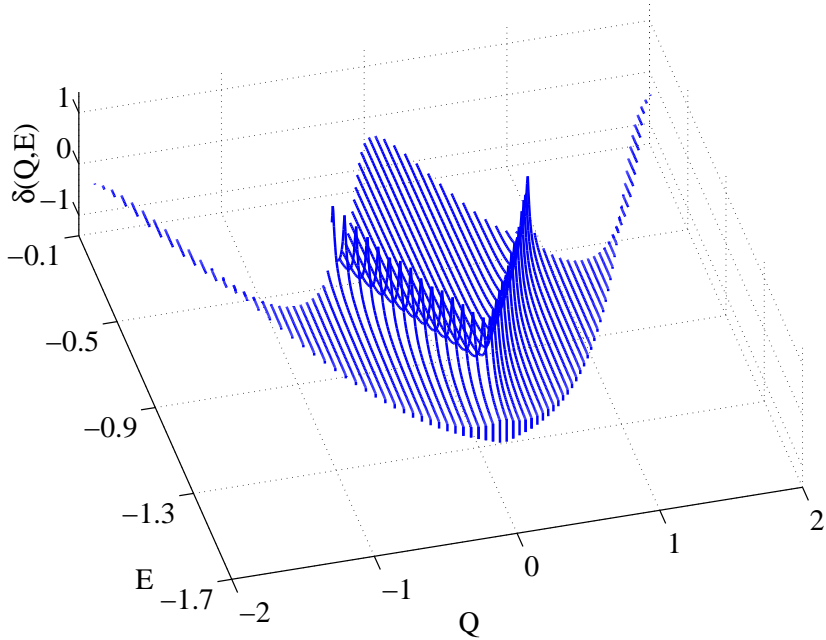}
  \caption{Even scattering phase-shift $\delta_e(E,Q)$ of a two-particle
    Bloch-wave versus energy $E$ and momentum $Q$. With a less deeply bound
    state the resonances remain above the lower band edge. Parameter: $u=1.5$,
    $v=-5$, $g=0.03$.}
  \label{Kmat2}
\end{figure}

For an experimental verification of the relative atomic momentum dependence of
the Feshbach resonance, as indicated in Figs.~\ref{Kmat} and \ref{Kmat2}, one
could study such scattering resonances for a variable filling of the lowest
energy band in the lattice. In this way, due to the Pauli exclusion principle,
for an increased band filling different regions of the relative atomic
quasi-momentum are subsequently filled up. The variation of the lineshape of
the Feshbach resonance is determined by the collisional properties of atoms in
the lattice.  For example, if we take the situation of Fig.~\ref{Kmat}, for
small atom filling the scattering resonance would disappear. The Feshbach
resonance here could only be observed for a large atom filling.
\section{Conclusions and Outlook}
We have considered Feshbach scattering of two Fermions in a one-dimensional
optical lattice. Due to the reduction of translation invariance, the relative
and center-of-mass motion are coupled and we present the scattering theory in
the crystal momentum basis. Within a single band approximation, we have
calculated numerically Feshbach resonances and demonstrate that the position
of the Feshbach resonance depends selectively on the center-of-mass momentum,
due to the non-parabolic shape of the energy band. The simple resonance
structure suggests that a semi-analytical calculation based on an
effective-mass approximation should be applicable and is currently pursued.

In the present article, we have only considered Feshbach resonances that are
much weaker than the interband separation. Thus, we could limit the discussion
to a single energy band.  Wider Feshbach resonances will involve the coupling
of several bands. Going beyond the single band approximations will provide
interesting new physics as already advanced in Ref.~\cite{diener06} for the
case of potential scattering. This will be examined, based on the present
formulation, in a forthcoming publication.

\ack
We gratefully acknowledge financial support by the SFB/TRR 21 {\em ``Control
  of quantum correlations in tailored matter''} funded by the Deutsche
Forschungsgemeinschaft (DFG). M.~G. and R.~W. are also thankful for the
hospitality of the Centro International de Ciencias A.C. (CICC) in Cuernavaca,
Mexico.




\section*{References}
\bibliographystyle{unsrt}

\bibliography{MyPublications,bec,DocLit}

\end{document}